\documentclass[%
 aip,
 amsmath,amssymb,
 preprint,%
 reprint,%
]{revtex4-1}

\usepackage{graphicx}
\usepackage{dcolumn}
\usepackage{bm}
\usepackage{xcolor}
\usepackage{booktabs}
\usepackage{float}

\usepackage[utf8]{inputenc}
\usepackage[T1]{fontenc}
\usepackage{mathptmx}
\usepackage{lineno}

\begin{document}

\title[TorbeamNN: Machine learning-based steering of ECH mirrors on KSTAR]{TorbeamNN: Machine learning based steering of ECH mirrors on KSTAR}

\author{Andrew Rothstein}
\affiliation{Department of Mechanical and Aerospace Engineering, Princeton University, Princeton, NJ, US}
\author{Minseok Kim}
\affiliation{Department of Mechanical and Aerospace Engineering, Princeton University, Princeton, NJ, US}
\author{Minho Woo}
\affiliation{Korean Institute of Fusion Energy, Daejeon, South Korea}
\author{Minsoo Cha}
\affiliation{Department of Energy Systems Engineering, Seoul National University, Seoul, South Korea}
\author{Cheolsik Byun}
\affiliation{Department of Mechanical and Aerospace Engineering, Princeton University, Princeton, NJ, US}
\author{Sangkyeun Kim}
\affiliation{Princeton Plasma Physics Laboratory, Princeton, NJ, US}
\author{Keith Erickson}
\affiliation{Princeton Plasma Physics Laboratory, Princeton, NJ, US}
\author{Youngho Lee}
\affiliation{Korean Institute of Fusion Energy, Daejeon, South Korea}
\author{Josh Josephy-Zack}
\affiliation{Princeton Plasma Physics Laboratory, Princeton, NJ, US}
\author{Jalal Butt}
\affiliation{Department of Mechanical and Aerospace Engineering, Princeton University, Princeton, NJ, US}
\author{Ricardo Shousha}
\affiliation{Princeton Plasma Physics Laboratory, Princeton, NJ, US}
\author{Mi Joung}
\affiliation{Korean Institute of Fusion Energy, Daejeon, South Korea}
\author{June-Woo Juhn}
\affiliation{Korean Institute of Fusion Energy, Daejeon, South Korea}
\author{Kyu Dong Lee}
\affiliation{Korean Institute of Fusion Energy, Daejeon, South Korea}
\author{Egemen Kolemen}\email{ekolemen@pppl.gov}
\affiliation{Department of Mechanical and Aerospace Engineering, Princeton University, Princeton, NJ, US}
\affiliation{Princeton Plasma Physics Laboratory, Princeton, NJ, US}

\begin{abstract}
    We have developed TorbeamNN: a machine learning surrogate model for the TORBEAM ray tracing code to predict electron cyclotron heating and current drive locations in tokamak plasmas. TorbeamNN provides more than a 100 times speed-up compared to the highly optimized and simplified real-time implementation of TORBEAM without any reduction in accuracy compared to the offline, full fidelity TORBEAM code. The model was trained using KSTAR electron cyclotron heating (ECH) mirror geometries and works for both O-mode and X-mode absorption. The TorbeamNN predictions have been validated both offline and real-time in experiment. TorbeamNN has been utilized to track an ECH absorption vertical position target in dynamic KSTAR plasmas as well as under varying toroidal mirror angles and with a minimal average tracking error of 0.5cm. 
\end{abstract}

\maketitle

\section{Introduction}\label{sec:intro}
Radio frequency (RF) heating sources are planned as primary heating sources for future tokamak reactors owing to their reliability and non-inductive heating power. ECH is the primary method planned for ITER ~\cite{omori_overview_2011}, and is a flexible actuator that provides localized heating to electrons and noninductive current drive through electron cyclotron current drive (ECCD). ECH and ECCD have a large number of applications in tokamak plasma control from NTM suppression with ECCD ~\cite{kolemen_state---art_2014,park_initial_2024,zohm_control_2007,bardoczi_direct_2023,maraschek_active_2005}, edge ECCD for improved ELM suppression access \cite{logan_access_2023,hu_effects_2024}, impurity  shielding effects \cite{sertoli_interplay_2015}, density pump out \cite{wang_understanding_2017}, and scenario development \cite{holcomb_steady_2014}. The common requirement in all of these applications is accurate aiming of ECH to achieve the desired goal. 

As ECH involves injecting RF waves into the tokamak plasma, the path of the RF wave is deflected by interactions with the plasma density and magnetic fields. The absorption location is determined by the toroidal magnetic field where the radio wave frequency matches electron cyclotron frequency. Additionally, the current drive efficiency, $\eta_{CD}$, changes with $T_e$ as $\eta_{CD}\propto T_e/n_e$. While some specific ECH tasks such as NTM control can be achieved without real-time ray-tracing codes \cite{hennen_real-time_2010,berrino_automatic_2006}, a flexible real-time control system capable of capable of using ECH for the various tasks possible in a fusion power plant will, with present control research, require a real-time ray-tracing code and must have information of the real-time plasma conditions, namely the magnetic field and electron density and temperature, in order to find the absorption location of the ECH waves. 

TORBEAM is a ray-tracing code that provides a solution to this problem and can be run in a reduced form real-time in approximately 10ms \cite{poli_torbeam_2018}. This physics-based code was used in experiments on ASDEX \cite{reich_real-time_2015} and DIII-D \cite{kolemen_state---art_2014}. However, running the full TORBEAM code in real-time requires full magnetic field information as well as $n_e$ and $T_e$ profiles, where tokamaks such as KSTAR that do not presently have a real-time $T_e$ profile. Additionally, the 10ms computation time was deemed acceptable for previous applications, but it introduces a 10ms delay to any control, which is not desirable for a plasma control system (PCS) with fast transients present, such as ELMs or sawteeth with  periods on the order of 20-50ms \cite{ahn_confinement_2012,jeong_demonstration_2015}. Since these transients change the plasma profiles and equilbria, they change the trajectory of the ECH rays at time scales that real-time TORBEAM cannot capture, while the faster TorbeamNN surrogate model can. It should be noted that at faster ELM periods below 10ms, it is not possible for the ECH hardware to keep up with the dynamic plasma conditions. 

A key motivation for quicker ECH ray-tracing is the future goal of multi-tasking ECH, such as the planned ECH control system on ITER, where ECH will be used for multiple tasks within a single shot. One of the fastest and most dangerous things controlled with ECH is NTM control. These instabilities can grow within 10s of ms to dangerous sizes. While real-time TORBEAM has been proven to be appropriate when focusing on just NTM suppression, it can be reasonably assumed that faster calculation will be needed to increase steering speeds if we begin adding additional control objectives such as scenario optimization or sawtooth pacing. Additional time saved allows both for more time to make decisions and allocate ECH power to individual tasks as well as steer mirrors to desired task locations. 

Machine learning (ML) surrogate models have been proven to be useful for control in fusion plasmas \cite{boyer_real-time_2019,morosohk_accelerated_2021,morosohk_neural_2021,rothstein_initial_2024} and utilizing surrogate models has the potential to speed up computation time to provide better tracking and faster response times. Additionally, surrogate models can leverage other sources of similar information, such as acquiring information about $T_e$ from plasma pressure and stored energy. ML surrogate models can also be trained on the full version of the physics-based code without the need to use a reduced real-time capable version of the code. 

In this paper we introduce TorbeamNN: a ML surrogate model for the TORBEAM code. This surrogate model provides the maximum absorption location of the ECH rays in the plasma. This model has been trained, deployed, and experimentally tested on the KSTAR tokamak in H-mode plasmas. 

The rest of the paper is organized as follows: In Section \ref{sec:model} we present the offline training and validation of the TorbeamNN models for KSTAR, then in Section \ref{sec:exp} we show experimental results of model deployment and control on KSTAR, and finally in Section \ref{sec:conc} we give concluding remarks and present future directions for application on KSTAR and other machines. 

\section{Model Training}\label{sec:model}

\subsection{Dataset Gathering}
There are presently three KSTAR gyrotrons that operate at a frequency of 105GHz, named EC2, EC4, and EC5. Each is able to operate in both X-mode and O-mode operation. A total of 6 TorbeamNN models were trained for each gyrotron launcher geometry and mode operation. The decision to have separate models was made to improve the learning of the surrogate models as X-mode versus O-mode ECH absorption has sufficiently different physics that a small surrogate model likely cannot capture. 

The final training set used KSTAR experimental data from the 2024 campaign with the tungsten divertor. A total of 2712 unique time slices were used from 336 different shots. Random gyrotron angles were selected within the safe operating space for each gyrotron mirror and then the full-fidelity offline TORBEAM code was run on each plasma time slice and mirror angle. For each gyrotron and each operating mode, the TORBEAM calculation was run a total of 647,267 times to provide a large dataset to train the surrogate models and was split into a 80-10-10 training, testing, and validation split. 

\begin{figure}
    \centering
    \includegraphics[width=0.75\linewidth]{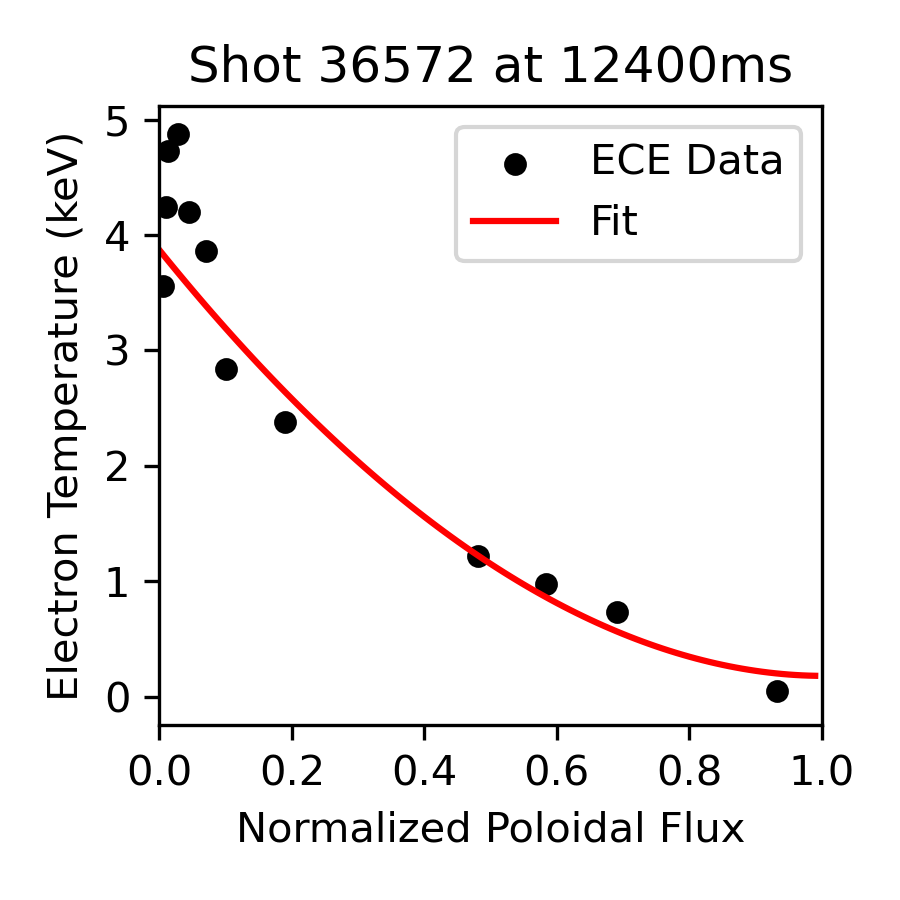}
    \caption{$T_e$ values measured from the ECE diagnostic (black) along with a line of best fit using a 3 parameter function (red). }
    \label{fig:Te}
\end{figure}

This dataset was selected to cover the whole range of operation of plasma shapes and $B_T$ while covering the space of achievable electron density and temperature profiles. For offline TORBEAM calculations, the necessary electron temperature profiles were generate with simple fits to the electron cyclotron emission data \cite{jeong_electron_2010} shown in Fig. \ref{fig:Te}. A simple 3 parameter function was assumed with the form 
$$T_e(\psi_N)=a(1-\psi_N)^\alpha+b$$
The electron density profiles were generated by a separate ML surrogate model using $CO_2$ interferometer data \cite{kim_design_2024,juhn_multi-chord_2021}. 

\subsection{Model Architecture and Performance}
The final set of inputs chosen for TorbeamNN are shown in Table \ref{tab:torbeamnn}. All inputs come from hardware readings from the ECH mirrors or EFIRT1 \cite{ferron_real_1998} except for the real-time $n_e$ profile estimate. This value is calculated by a ML surrogate model based off of the $CO_2$ interferometry data \cite{kim_design_2024}. While a full $n_e$ profile is provided by the PCS algorithm, only three points were necessary for the model to have accurate predictions: one value at the core, one off-axis ($\psi_N=0.5$), and one at the end ($\psi_N=1.0$). The minor radius is defined as $(R_{max}-R_{min})/2$ where $R_{max}$ is the major radius value of the low field side last closed flux surface and $R_{min}$ is the same on the high field side. 

Initially, more features were included as input to the surrogate model, such as inner and outer flux surface locations, triangularity, gap to various walls, and $W_{MHD}$. However, in training it was found these parameters could be removed without reducing the $R^2$ values and thus contained redundant information. This process was also how the $n_e$ profile was found to just require the $3$ points. 

\begin{table*}[t]
    \small
    \centering
    {
        \begin{tabular}{|ccc|cc|}
        \hline
        \textbf{Inputs} & \textbf{Description}  & \textbf{Source} & \textbf{Train Min} & \textbf{Train Max} \\
        \hline
        $\theta$ & EC mirror poloidal angle & ECH hardware & $8.02^\circ$ & $28.50^\circ$ \\
        $\varphi$ & EC mirror toroidal angle & ECH hardware & $-22.89^\circ$ & $22.89^\circ$\\
        $B_T$ (T) & Vacuum toroidal field & rtEFIT & $-2.49$ & $-1.59$ \\
        $I_P$ (MA) & Plasma current & rtEFIT & $-0.71$ & $-0.38$ \\
        $R_0$ (m) & Magnetic axis R location & rtEFIT & $1.75$ & $1.90$ \\
        $Z_0$ (m) & Magnetic axis Z location & rtEFIT & $-0.10$ & $0.09$ \\
        $a$ (m) & Minor radius & rtEFIT & $0.44$ & $0.50$ \\
        $\beta_N$ & Normalized pressure & rtEFIT & $0.17$ & $3.09$  \\
        $\kappa$ & Elongation & rtEFIT & $1.47$ & $2.15$ \\
        $l_i$ & Plasma inductance & rtEFIT & $0.80$ & $1.86$ \\
        $V$ $(m^{-3})$ & Plasma volume & rtEFIT & $10.15$ & $14.70$ \\
        $n_e$ $(10^{19}/m^3)$ & Electron density & $n_e$ RECON & $0.70$ & $8.91$ \\
        \hline 
        \textbf{Outputs} & \textbf{Description} & \textbf{Source} & \textbf{Train Min} & \textbf{Train Max}\\
        \hline 
        $\rho_{pol}$ & Poloidal rho location & TorbeamNN & $0.00$ & $0.96$ \\
        $R$ (m) & R absorption location & TorbeamNN & $1.32$ & $2.22$\\
        $Z$ (m) & Z absorption location & TorbeamNN & $-0.76$ & $0.66$ \\
        $\eta_{CD}$ & CD efficiency & TorbeamNN & $-0.31$ & $0.31$ \\
        \hline
        \end{tabular}
    }
    \caption{Input and output signals for TorbeamNN model. rtEFIT refers to the fast loop real-time EFIT without MSE constraint, also known as EFITRT1. The final 2 columns give the range of parameters used in training for TorbeamNN with the Train Min being the minimum value in the training set and Train Max being the maximum. This gives the full range of each parameter, although the majority of the training set lies closer to the mean. For electron density, just the values of the core $n_e$ are shown.}
    \label{tab:torbeamnn}
\end{table*}

In real-time, we assumed sufficient information about electron temperature would be provided by normalized plasma pressure, $\beta_N$, which was later verified by our successful prediction of ECCD efficiency shown later in this section. We also hypothesized that the global plasma shape parameters held sufficient information about the magnetic plasma equilibrium and the full 3D grid of magnetic field values used by TORBEAM was unnecessary, which again proved to be true as shown in our surrogate model performance metrics later this section. This allows us to skip the costly real-time computation of calculating $B_R$, $B_T$, $B_Z$, and $\rho$ at a full grid of points and can utilize quantities already computed by EFITRT1. 

The final models were trained using the Keras library with the Adam optimizer and a learning rate of 0.00063. The final architecture chosen has $3$ dense layers of $60$ neurons each and uses the ReLU activation function for each dense layer followed by a linear activation for the final layer. Other model sizes were explored, but larger models did not significantly improve training performance and this architecture was small enough to reach inference time goals described later. Additional model size reduction could be performed to further reduce inference times. 

\begin{figure*}
    \centering
    \includegraphics[width=0.9\linewidth]{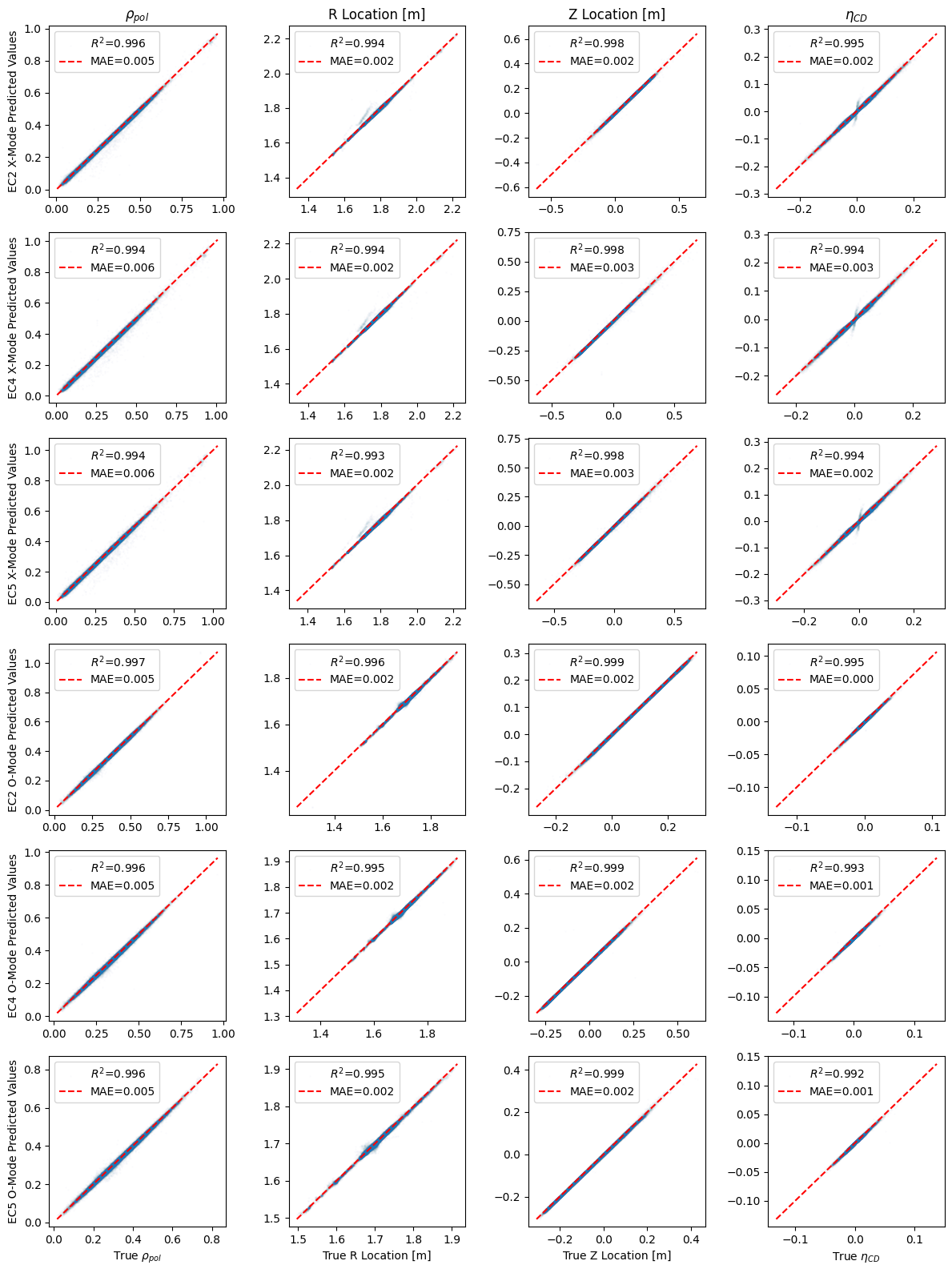}
    \caption{$R^2$ and mean absolute error (MAE) values for TorbeamNN training. From left to right: the columns show $\rho_{pol}$, $R$ absorption location, $Z$ absorption location, and $\eta_{CD}$. From top to bottom, the row show EC2, EC4, then EC5 predictions in X-mode followed by those three gyrotrons in O-mode. }
    \label{fig:r2}
\end{figure*}

The offline model performance on the test dataset is shown in Fig. \ref{fig:r2}. All $R^2$ values are above $0.992$ and the largest mean absolute error in $Z$ location prediction is $0.2$cm, or in other words we can expect our $Z$ absorption location predictions to be on average within $0.2$cm of the true location. These absorption locations correspond to the second harmonic absorption location, as this is the primary absorption location in typical KSTAR plasmas for $105$GHz ECH \cite{joung_design_2024}. Qualitatively, O-mode appears to generally do better than X-mode, which makes physical sense as there will be less deflection compared to the vacuum calculation. While the O-mode absorption is typically less than X-mode, approximately only 50-80\% may be absorbed depending on the conditions, the location returned by TorbeamNN is the maximum deposition location. Also of minor note is that TorbeamNN seems to have worse predictions of $\eta_{CD}$ when the current drive is near $0$, corresponding to perpendicular injection for ECH-only heating, as seen in the small vertical spread in predicted $\eta_{CD}$ around true $\eta_{EC}=0$. 

\section{Experimental Deployment}\label{sec:exp}

\begin{figure}
    \centering
    \includegraphics[width=0.5\linewidth]{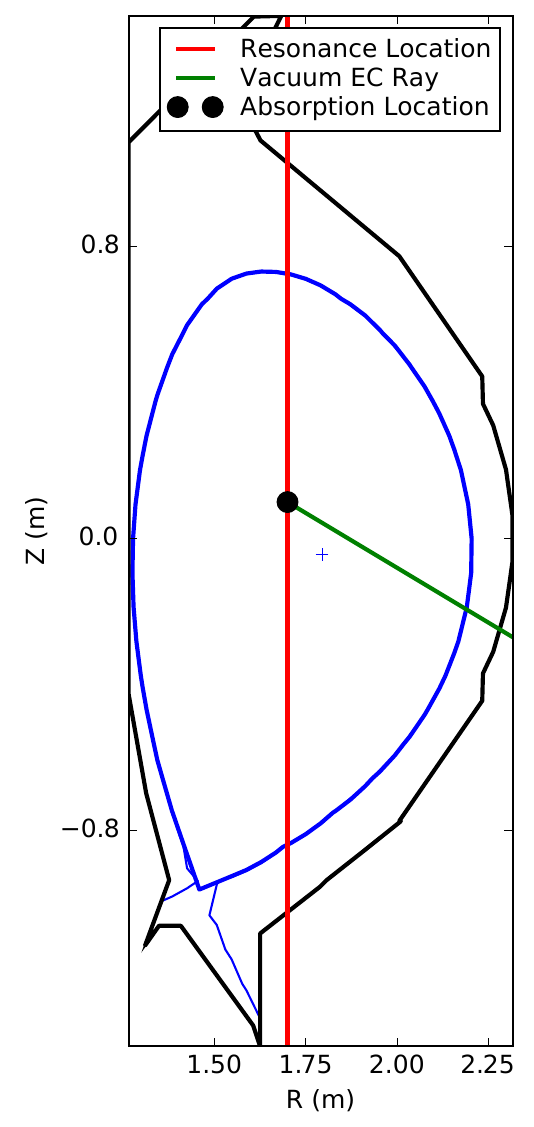}
    \caption{Sketch of EC vacuum geometry. The vertical red line is the the resonance based on $B_T$ and gives $R$. The green line is the geometric EC ray that is a straight line with no deflection. The black dot is located at $(R,Z_{vac})$ to give vacuum absorption $Z_{vac}$. }
    \label{fig:vacuum_EC}
\end{figure}

ECH steering control on KSTAR is done with respect to the vacuum absorption $Z_{vac}$ location. To find this location, first the toroidal magnetic field is used to find the radial resonant absorption location $R$ where the EC frequency matches the electron gyrofrequency. With the $R$ location of absorption, the ECH mirror geometry, and ECH mirror angles, we assume no refraction in a vacuum and so the EC wave travels in a straight line from the launcher mirror to the radial absorption location to give us the $Z_{vac}$ absorption location as seen in Fig. \ref{fig:vacuum_EC}. $Z=0$ is defined as the horizontal axis across the center of the vacuum vessel. 

For most control applications, it is desirable to have a control target of $\rho_{pol}$ rather than a Z position. However, for initial testing and following KSTAR conventions, $Z_{vac}$ was used as the control target. While the KSTAR PCS can convert $Z_{vac}$ to $\rho_{pol}$, due to limited experimental time this capability was not able to be tested.  

\subsection{PCS Development and Tuning}
The TorbeamNN Keras models developed offline were transformed into the PCS compatible C functions with the keras2c library \cite{conlin_keras2c_2021}. After this conversion and implementation on the KSTAR PCS, each surrogate model was timed and runs consistently under $60\mu s$ per model with typical run times closer to $50\mu s$. This represents a speed-up of around 100 times compared to the real-time TORBEAM code. In total, the full CPU cycle time for all three gyrotrons run in series was always under $180\mu s$. Further speed-up was not needed, but these models could easily be run in parallel for further computational speed-up.  

When using the real-time TORBEAM code that takes around 10-20ms, TORBEAM is the bottle-neck for faster EC steering compare to the 1-5ms time scale for real-time EFIT and below $0.5$ms for the real-time $n_e$ reconstruction. At maximum speed, the ECH mirror can move around $0.4^\circ$ in 10ms, not accounting for delays in receiving the command. There is an additional $\approx 20ms$ delay between when a poloidal angle command is received to when the mirror begins to move, which is effectively doubled if  real-time TORBEAM is used for the ray-tracing calculation because first the ray must be calculated before the command can be sent. 

\begin{figure}
    \centering
    \includegraphics[width=\linewidth]{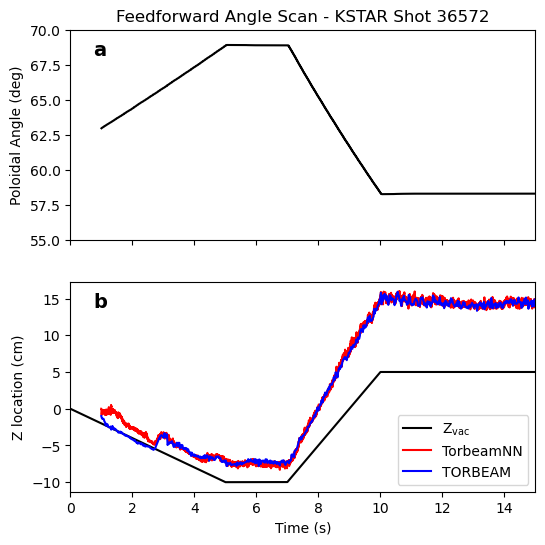}
    \caption{Results of feedforward poloidal angle scan using EC5. \textbf{a)} Poloidal angle over time. \textbf{b)} Z locations of: vacuum calculation for ECH deposition (black), PCS results of TorbeamNN (red), and offline calculation using full-fidelity TORBEAM (blue). }
    \label{fig:ff}
\end{figure}

During the experiments, only EC5 feedback steering was operational and all results shown here are for EC5 only. The first testing was done with the gyrotron poloidal angles set in feedforward control with fixed toroidal angles. Shown in Fig. \ref{fig:ff} are the results of feedforward steering and the TorbeamNN predictions along with the offline validation EC absorption location using TORBEAM. Across the whole discharge, when comparing TorbeamNN versus offline TORBEAM we achieve an $R^2$ value of $0.995$ and a mean absolute error of $0.44$cm. An important discrepancy is in the dynamic ramp-up portion of the shot prior to the LH transition around 3 seconds where the model has significantly worse predictions. This is caused by inaccuracies in real-time versus offline real-time EFITs as the offline magnetic sensor settings were not accurately updated for this campaign. Here, we have found the post-experiment TORBEAM results are inaccurate due to incorrect magnetic equilibria. Thus, the results from the early phase are inaccurate and should not be considered. Only looking at the flattop portion of the shot after 3 seconds, we have an $R^2$ of $0.998$ and mean absolute error of $0.28$cm. 

Compared to both the accuracy of the online and offline inputs, this accuracy is acceptable. The main sources of error come from real-time EFIT and the real-time $n_e$ reconstruction because the ECH hardware provides accurate angle steering well below $1^\circ$. The accuracy of the real-time EFIT inputs is explored in \cite{ferron_real_1998} and for parameters such as the boundary location, the average error is around $0.01-1$cm with the worst errors being $1-4$cm. Our mean absolute error of $0.28$cm is reasonably within the error range of real-time EFIT. For the accuracy of the $n_e$ reconstruction, we look to \cite{kim_design_2024} where the mean squared error was found to be $1.31 \times 10^{17} m^{-3}$ using data from the 2023 KSTAR campaign. Without further information it is difficult to quantify the errors precisely, but they appear to be under 10\%. Between the errors from the real-time EFIT and real-time $n_e$, our surrogate model provides results on the order of the accuracy of the input values. 

\subsection{TorbeamNN control results}
\begin{figure}
    \centering
    \includegraphics[width=\linewidth]{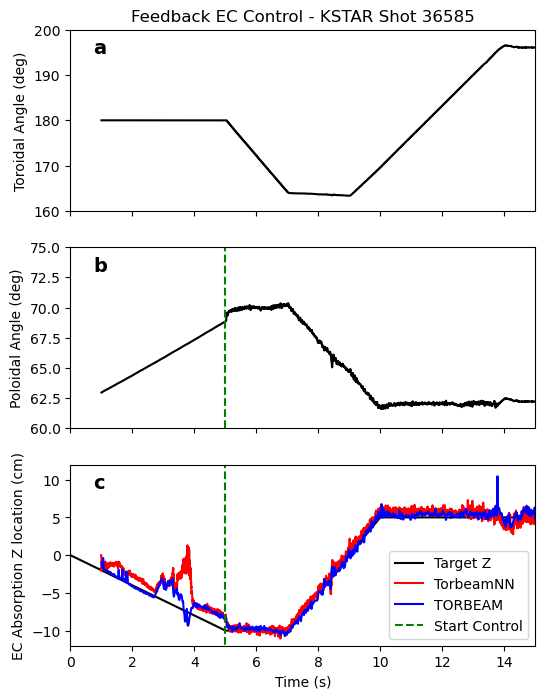}
    \caption{Results from feedback control of EC5 mirror. \textbf{a)} Feedforward toroidal angle scan values. \textbf{b)} Poloidal angle of EC5 over time. Feedback control starts at 5 seconds (vertical green line) and continues for remainder of shot. \textbf{c)} Z locations for: target Z value (black), real-time TorbeamNN calculated absorption location (red), and offline calculated absorption location (blue). Feedback control starts at 5 seconds (vertical green line).}
    \label{fig:exp}
\end{figure}

With feedforward validation complete, standard PID tuning was done to find optimal gains for the vacuum $Z$ target tracking. A standard PID controller was implemented of the form
$$u(t)=K_pe(t)+K_i\int e(t)\mathrm{d}t+K_d\frac{\mathrm de}{\mathrm dt}$$
Where $u(t)$ is our actuator of the EC mirror angle, $e(t)$ is the error between target and actual $Z$ locations, and $K_p,K_i,K_d$ represent our proportional, integral, and derivative gains, respectively. Only the proportional and integral gain terms were found to be necessary to get a quick response and minimal steady state error. We achieved successful target Z location tracking as seen in Fig. \ref{fig:exp} during a dynamic plasma shot while simultaneously changing the toroidal mirror angle. We see similar poor performance of TorbeamNN prior to the LH transition around 3 seconds, so feedback control was chosen to start at 5 seconds after the dynamic ramp-up portion of the shot has completed. 

During the control shot shown in Fig. \ref{fig:exp}, the mean absolute error is 0.535cm over the interval from 5.1 seconds to 15 seconds, allowing the PID controller 100ms to ramp-up after initializing at 5 seconds. This average error is within the expected range of fluctuations due to changes in density caused by transients present in the plasma and is sufficiently accurate for any application of real-time ECH or ECCD control. Additionally, the $K_i$ term could be further optimized to reduce steady state error, as the Classical Ziegler–Nichols used to tune $K_i/K_p$ is known to underestimate the $K_i$ term. Future experimental time can be used to further reduce this tracking error.

\section{Conclusion}\label{sec:conc}
TorbeamNN was trained offline covering the operational space of the 2024 KSTAR campaign over the full range of allowable ECH mirror angles. Along with a well-tuned PID controller, the TorbeamNN provides the information to quickly steer and track vacuum field $Z$ locations. Similarly, $\rho$ tracking should be achievable by converting $\rho$ targets to $Z$ targets in real-time, however there was not sufficient experimental time to verify this. 

In the future when KSTAR's real-time MSE diagnostic is available, real-time MSE-constrained EFITs will be used for real-time q-profile tracking for applications in NTM control. Other applications such as tungsten shielding or edge ECCD for easier ELM suppression access can utilize this q-profile tracking capability as well. 

Future real-time controllers could take advantage of the speed and differentiability of ML surrogate models to avoid PID control altogether. A brute force option could leverage TorbeamNN's significantly faster cycle time to repeatedly search for the optimal poloidal angle to reach the desired target. A more sophisticated approach could leverage a dense neural network's ability to be differentiated directly to compute the change in absorption position based on the change in poloidal angle. Either approach will provide faster convergence of the ECH mirror steering. 

A further useful extension of TorbeamNN would be predicting the full ECH and ECCD profiles in real-time. The best suited reactors for this task would be a well-diagnosed machine with accurate real-time equilibria and full $n_e$ and $T_e$ profiles to provide accurate information not just on absorption location, but also reliable information on heating and current drive efficiencies. Having access to the full ECH and ECCD deposition profiles would enable advanced real-time control capabilities such as optimizing ECCD profiles to provide flat, broad ECCD profiles in changing plasma conditions or tracking maximally efficient ECCD locations in real-time. 

Most future tokamaks plan for various forms of RF heating as it is a reliable, non-inductive heating source. Utilizing ML surrogate models in place of a real-time ray-tracing code provides RF absorption locations more than 100 times faster than physics-based codes without losing accuracy. Future reactors' plasma control systems should leverage this capability to have rapid, accurate knowledge about how their RF heating sources are affecting the plasma to make better-informed control decisions. 

\section*{Acknowledgments}

This material is based upon work supported by the U.S. Department of Energy, Office of Science, Office of Fusion Energy Sciences, using the DIII-D National Fusion Facility, a DOE Office of Science user facility, under Award DE-FC02-04ER54698. Additionally, this material is supported by the National Science Foundation Graduate Research Fellowship under Grant No. DGE-2039656 and by the U.S. Department of Energy, under Awards DE-SC0015480. 

\section*{Disclaimer}

This report was prepared as an account of work sponsored by an agency of the United States Government. Neither the United States Government nor any agency thereof, nor any of their employees, makes any warranty, express or implied, or assumes any legal liability or responsibility for the accuracy, completeness, or usefulness of any information, apparatus, product, or process disclosed, or represents that its use would not infringe privately owned rights. Reference herein to any specific commercial product, process, or service by trade name, trademark, manufacturer, or otherwise does not necessarily constitute or imply its endorsement, recommendation, or favoring by the United States Government or any agency thereof. The views and opinions of authors expressed herein do not necessarily state or reflect those of the United States Government or any agency thereof.

\bibliography{KSTAR-TorbeamNN.bib}

\begin{thebibliography}{10}

\bibitem{omori_overview_2011}
T.~Omori, M.A. Henderson, F.~Albajar, S.~Alberti, U.~Baruah, T.S. Bigelow, B.~Beckett, R.~Bertizzolo, T.~Bonicelli, A.~Bruschi, J.B. Caughman, R.~Chavan, S.~Cirant, A.~Collazos, D.~Cox, C.~Darbos, M.R. De~Baar, G.~Denisov, D.~Farina, F.~Gandini, T.~Gassmann, T.P. Goodman, R.~Heidinger, J.P. Hogge, S.~Illy, O.~Jean, J.~Jin, K.~Kajiwara, W.~Kasparek, A.~Kasugai, S.~Kern, N.~Kobayashi, H.~Kumric, J.D. Landis, A.~Moro, C.~Nazare, Y.~Oda, I.~Pagonakis, B.~Piosczyk, P.~Platania, B.~Plaum, E.~Poli, L.~Porte, D.~Purohit, G.~Ramponi, S.L. Rao, D.A. Rasmussen, D.M.S. Ronden, T.~Rzesnicki, G.~Saibene, K.~Sakamoto, F.~Sanchez, T.~Scherer, M.A. Shapiro, C.~Sozzi, P.~Spaeh, D.~Strauss, O.~Sauter, K.~Takahashi, R.J. Temkin, M.~Thumm, M.Q. Tran, V.S. Udintsev, and H.~Zohm.
\newblock Overview of the {ITER} {EC} {H}\&{CD} system and its capabilities.
\newblock {\em Fusion Engineering and Design}, 86(6-8):951--954, October 2011.

\bibitem{kolemen_state---art_2014}
E.~Kolemen, A.S. Welander, R.J. La~Haye, N.W. Eidietis, D.A. Humphreys, J.~Lohr, V.~Noraky, B.G. Penaflor, R.~Prater, and F.~Turco.
\newblock State-of-the-art neoclassical tearing mode control in {DIII}-{D} using real-time steerable electron cyclotron current drive launchers.
\newblock {\em Nuclear Fusion}, 54(7):073020, July 2014.

\bibitem{park_initial_2024}
Y~S Park, M~H Woo, S~A Sabbagh, H~S Han, B~H Park, J~S Kang, and H~S Kim.
\newblock Initial results from neoclassical tearing mode stabilization experiment in {KSTAR} high normalized beta plasmas.
\newblock {\em Plasma Physics and Controlled Fusion}, 66(12):125013, December 2024.

\bibitem{zohm_control_2007}
H~Zohm, G~Gantenbein, F~Leuterer, M~Maraschek, E~Poli, L~Urso, and {the ASDEX Upgrade Team}.
\newblock Control of {NTMs} by {ECCD} on {ASDEX} {Upgrade} in view of {ITER} application.
\newblock {\em Plasma Physics and Controlled Fusion}, 49(12B):B341--B347, December 2007.

\bibitem{bardoczi_direct_2023}
L.~Bardóczi, R.J. La~Haye, E.J. Strait, N.C. Logan, S.P. Smith, N.J. Richner, and J.D. Callen.
\newblock Direct preemptive stabilization of m,n=2,1 neoclassical tearing modes by electron cyclotron current drive in the {DIII}-{D} low-torque {ITER} baseline scenario.
\newblock {\em Nuclear Fusion}, 63(9):096021, September 2023.

\bibitem{maraschek_active_2005}
M~Maraschek, G~Gantenbein, T.P Goodman, S~Günter, D.F Howell, F~Leuterer, A~Mück, O~Sauter, H~Zohm, Contributors To The Efda-Jet Workprogramme, and The Asdex~Upgrade Team.
\newblock Active control of {MHD} instabilities by {ECCD} in {ASDEX} {Upgrade}.
\newblock {\em Nuclear Fusion}, 45(11):1369--1376, November 2005.

\bibitem{logan_access_2023}
Nikolas Logan, Brendan~C. Lyons, M~Knölker, Qiming Hu, Tyler Cote, and Philip~B Snyder.
\newblock Access to stable, high pressure tokamak pedestals using local electron cyclotron current drive.
\newblock {\em Nuclear Fusion}, November 2023.

\bibitem{hu_effects_2024}
Qiming Hu, Nikolas Logan, Qingquan Yu, and Alessandro Bortolon.
\newblock Effects of edge-localized electron cyclotron current drive on edge-localized mode suppression by resonant magnetic perturbations in {DIII}-{D}.
\newblock {\em Nuclear Fusion}, February 2024.

\bibitem{sertoli_interplay_2015}
M.~Sertoli, T.~Odstrcil, C.~Angioni, and {the ASDEX Upgrade Team}.
\newblock Interplay between central {ECRH} and saturated ( \textit{m} , \textit{n} ) = (1, 1) {MHD} activity in mitigating tungsten accumulation at {ASDEX} {Upgrade}.
\newblock {\em Nuclear Fusion}, 55(11):113029, September 2015.

\bibitem{wang_understanding_2017}
X.~Wang, S.~Mordijck, E.J. Doyle, T.L. Rhodes, L.~Zeng, G.R. McKee, M.E. Austin, O.~Meneghini, G.M. Staebler, and S.P. Smith.
\newblock Understanding {ECH} density pump-out in {DIII}-{D} {H}-mode plasmas.
\newblock {\em Nuclear Fusion}, 57(11):116046, November 2017.

\bibitem{holcomb_steady_2014}
C.T. Holcomb, J.R. Ferron, T.C. Luce, T.W. Petrie, J.M. Park, F.~Turco, M.A. Van~Zeeland, M.~Okabayashi, C.T. Lasnier, J.M. Hanson, P.A. Politzter, Y.~In, A.W. Hyatt, R.J. La~Haye, and M.J. Lanctot.
\newblock Steady state scenario development with elevated minimum safety factor on {DIII}-{D}.
\newblock {\em Nuclear Fusion}, 54(9):093009, September 2014.

\bibitem{hennen_real-time_2010}
B~A Hennen, E~Westerhof, P~W J~M Nuij, J~W Oosterbeek, M~R De~Baar, W~A Bongers, A~Bürger, D~J Thoen, M~Steinbuch, and {the TEXTOR Team}.
\newblock Real-time control of tearing modes using a line-of-sight electron cyclotron emission diagnostic.
\newblock {\em Plasma Physics and Controlled Fusion}, 52(10):104006, October 2010.

\bibitem{berrino_automatic_2006}
J.~Berrino, S.~Cirant, F.~Gandini, G.~Granucci, E.~Lazzaro, F.~Jannone, P.~Smeulders, and G.~D'Antona.
\newblock Automatic real-time tracking and stabilization of magnetic islands in a {Tokamak} using temperature fluctuations and {ECW} power.
\newblock {\em IEEE Transactions on Nuclear Science}, 53(3):1009--1014, June 2006.

\bibitem{poli_torbeam_2018}
E.~Poli, A.~Bock, M.~Lochbrunner, O.~Maj, M.~Reich, A.~Snicker, A.~Stegmeir, F.~Volpe, N.~Bertelli, R.~Bilato, G.D. Conway, D.~Farina, F.~Felici, L.~Figini, R.~Fischer, C.~Galperti, T.~Happel, Y.R. Lin-Liu, N.B. Marushchenko, U.~Mszanowski, F.M. Poli, J.~Stober, E.~Westerhof, R.~Zille, A.G. Peeters, and G.V. Pereverzev.
\newblock {TORBEAM} 2.0, a paraxial beam tracing code for electron-cyclotron beams in fusion plasmas for extended physics applications.
\newblock {\em Computer Physics Communications}, 225:36--46, April 2018.

\bibitem{reich_real-time_2015}
M.~Reich, R.~Bilato, U.~Mszanowski, E.~Poli, C.~Rapson, J.~Stober, F.~Volpe, and R.~Zille.
\newblock Real-time beam tracing for control of the deposition location of electron cyclotron waves.
\newblock {\em Fusion Engineering and Design}, 100:73--80, November 2015.

\bibitem{ahn_confinement_2012}
J-W. Ahn, H.-S. Kim, Y.S. Park, L.~Terzolo, W.H. Ko, J.-K. Park, A.C. England, S.W. Yoon, Y.M. Jeon, S.A. Sabbagh, Y.S. Bae, J.G. Bak, S.H. Hahn, D.L. Hillis, J.~Kim, W.C. Kim, J.G. Kwak, K.D. Lee, Y.S. Na, Y.U. Nam, Y.K. Oh, and S.I. Park.
\newblock Confinement and {ELM} characteristics of {H}-mode plasmas in {KSTAR}.
\newblock {\em Nuclear Fusion}, 52(11):114001, November 2012.

\bibitem{jeong_demonstration_2015}
J.~H. Jeong, Y.~S. Bae, M.~Joung, D.~Kim, T.~P. Goodman, O.~Sauter, K.~Sakamoto, K.~Kajiwara, Y.~Oda, J.~G. Kwak, W.~Namkung, M.~H. Cho, H.~Park, J.~Hosea, and R.~Ellis.
\newblock Demonstration of sawtooth period control with {EC} waves in {KSTAR} plasma.
\newblock {\em EPJ Web of Conferences}, 87:02016, 2015.

\bibitem{boyer_real-time_2019}
M.D. Boyer, S.~Kaye, and K.~Erickson.
\newblock Real-time capable modeling of neutral beam injection on {NSTX}-{U} using neural networks.
\newblock {\em Nuclear Fusion}, 59(5):056008, May 2019.

\bibitem{morosohk_accelerated_2021}
Shira~M. Morosohk, Mark~D. Boyer, and Eugenio Schuster.
\newblock Accelerated version of {NUBEAM} capabilities in {DIII}-{D} using neural networks.
\newblock {\em Fusion Engineering and Design}, 163:112125, February 2021.

\bibitem{morosohk_neural_2021}
S.M. Morosohk, A.~Pajares, T.~Rafiq, and E.~Schuster.
\newblock Neural network model of the multi-mode anomalous transport module for accelerated transport simulations.
\newblock {\em Nuclear Fusion}, 61(10):106040, October 2021.

\bibitem{rothstein_initial_2024}
Andrew Rothstein, Azarakhsh Jalalvand, Joseph Abbate, Keith Erickson, and Egemen Kolemen.
\newblock Initial testing of {Alfvén} eigenmode feedback control with machine-learning observers on {DIII}-{D}.
\newblock {\em Nuclear Fusion}, 64(9):096020, September 2024.

\bibitem{jeong_electron_2010}
S.~H. Jeong, K.~D. Lee, Y.~Kogi, K.~Kawahata, Y.~Nagayama, A.~Mase, and M.~Kwon.
\newblock Electron cyclotron emission diagnostics on {KSTAR} tokamak.
\newblock {\em Review of Scientific Instruments}, 81(10):10D922, October 2010.

\bibitem{kim_design_2024}
M.S. Kim, S.K. Kim, K.~Erickson, A.~Rothstein, R~Shousha, S.H Han, J.~Juhn, B.S. Kim, C.S Byun, J.~Butt, S.M. Yang, Q.~Hu, D.~Eldon, H.S. Han, N.~Logan, A.~Jalalvand, and E.~Kolemen.
\newblock Design of a pid controller for the pedestal top electron density at kstar.
\newblock In {\em Proceedings of the Annual APS-DPP Meeting}, Atlanta, GA, October 2024. APS-DPP.

\bibitem{juhn_multi-chord_2021}
June-Woo Juhn, K.~C. Lee, T.~G. Lee, H.~M. Wi, Y.~S. Kim, S.~H. Hahn, and Y.~U. Nam.
\newblock Multi-chord {IR}–visible two-color interferometer on {KSTAR}.
\newblock {\em Review of Scientific Instruments}, 92(4):043559, April 2021.

\bibitem{ferron_real_1998}
J.R Ferron, M.L Walker, L.L Lao, H.E.~St John, D.A Humphreys, and J.A Leuer.
\newblock Real time equilibrium reconstruction for tokamak discharge control.
\newblock {\em Nuclear Fusion}, 38(7):1055--1066, July 1998.

\bibitem{joung_design_2024}
Mi~Joung, Sonjong Wang, Sunggug Kim, Jongwon Han, Inhyuk Rhee, and Jonggu Kwak.
\newblock Design and operation results of {KSTAR} {ECH} system.
\newblock {\em Fusion Engineering and Design}, 203:114461, June 2024.

\bibitem{conlin_keras2c_2021}
Rory Conlin, Keith Erickson, Joseph Abbate, and Egemen Kolemen.
\newblock Keras2c: {A} library for converting {Keras} neural networks to real-time compatible {C}.
\newblock {\em Engineering Applications of Artificial Intelligence}, 100:104182, April 2021.

\end{thebibliography}

\end{document}